\documentclass[]{article}
\newtheorem{theorem}{Theorem}

\begin{document}

\title{Review of some classical gravitational superenergy tensors using computational techniques }
\author{
 A. Balfag\' on\thanks{Institut Qu\'\i mic de Sarri\` a, Laboratori de F\'\i sica Matem\` atica,Societat Catalana de F\'\i sica (I.E.C.)Universitat Ram\' on Llull e-mail: abalf@iqs.url.es} and X. Ja\' en.\thanks{Universitat Polit\` ecnica de Catalunya, Laboratori de F\'\i sica Matem\` atica, Societat Catalana de F\'\i sica (I.E.C.) e-mail: jaen@baldufa.upc.es}}
%
\maketitle
\begin{abstract}
We use computational algorithms recently developed by us to study  completely four index divergence free quadratic in Riemann tensor polynomials in GR. Some results are new and some other reproduce and/or correct known ones.
The algorithms are part of a {\it Mathematica} package called
{\it Tools of Tensor Calculus (TTC)}[web address:
http://baldufa.upc.es/ttc] 
\end{abstract}
\section{Introduction}
Recently the interest of tensors containing the interaction fields, and in particular the Riemann tensor, has been increased. The so-called superenergy tensors are good examples, having some particular properties \cite{Senovilla}. This interest is not new. The tensors of Bel-Robinson\cite{BelRobinson},Bel\cite{Bel}and Sachs\cite{Sachs} among others are some examples which have been studied in the past and having zero divergence in certain conditions. CD Collinson \cite{Collinson} has made a much deeper study giving all four index quadratic in Riemann and divergence free tensors in four dimension for any space-time. Nevertheless this kind of calculations are very hard to do by hand and some errors could arise. Having this in mind, recently, we have developed some computational algorithms able to handle index tensor  polynomial expressions.  The paradigm of these polynomials are those built with Riemann tensor as in the example

$$...\frac{1}{2}R_{i\,j}^{\,\,\,\,\,\,m\,n} \,R_{m\,n\,\,\,\,\,\,\,\,;l}^{\,\,\,\,\,\,\,\,\,k\,l}  + \,3\,R_{i\,m\,\,\,\,}^{\,\,\,\,\,\,\,\,k\,m} \,R_{j;l}^{\,\,\,\,\,\,\,\,\,l} \,\, ...$$

The implementation of the algorithms are part of a package called {\it TTC} \cite{TTC} ({\it Tools of Tensor Calculus}).  These algorithms allows us to perform a complete simplification of this kind of expressions. The simplification algorithms do not find the most beautiful expression but, besides zero result, a canonical one.  

Other authors who have worked on algorithmic simplification of index tensor polynomials are Ilyin and Kryuchov\cite{Ilyin}, Portugal\cite{Portugal}, Fulling et al.\cite{Fulling}.  Parker and Christiensen\cite{MathTensor} do not fall on this list because although they have a big package on tensor calculus, they don't work algorithmically in this topic but with libraries of rules. 
In a recent article\cite{XA1998} we  have solved the part of the simplification problem related to dummy indices and monoterm symmetry properties. Cyclic and in general multiterm symmetries properties were not considered. Also dimensional properties were left aside for further works, due to its complexity. 
Actually \cite{XACyclicRicci,XADim} we can handle multiterm and dimensional properties with the advantage that the algorithm includes the possibility to store a compacted version of the generated rules in order to increase functionality. This means that, although a first computation over a polynomial can be really slow, others on the same or on other sessions are faster than the first one. This allows us to {\it play} easily with some results by permuting indices or making some other operations. 

There are some trust  problems when using programs which do automatic computations, specially when no exact coindicences arise when compared with older papers. All we can say is that our algorithm has been tested, besides in the present paper, on a large set of known Riemann properties ( among others all  those included in \cite{MathTensor}) and calculations related to the Gauss-Bonnet invariant\cite{Christensen}. On the other hand the fact that the implementation of the algorithm is part of a package, {\it TTC},( from now on  we  will use the name  {\it TTC} to call both, the simplification algorithm and/or the package) it allows us to test the results on explicit space-times. We have tested the calculations on Vaydia metric confirming our results. 

The aim of this paper is:
\begin{itemize}
\item  To show the {\it TTC} capabilities on a real calculus. This can be of interest to researchers working on algebraic computation and/or working on superenergy tensors or on any other field where these techniques can be applied. 
\item To review some known results and correct some others. Specially we have corrected a minor part of the results of C.D. Collinson and some sign on the original Sachs paper \cite{Sachs}.
\item To prove some theorems about existence and uniquness of superenergy tensors.  
\end{itemize}

By superenergy tensors we understand, in this paper, four index quadratic in Riemann tensor having some chosen properties. Following the work of C.D. Collinson we will find divergence free tensors and from these we will analyze the index symmetry properties. We leave aside other properties such as the dominant super-energy property (DSEP) and others explained in reference\cite{Senovilla} 

The plan is as follows. In section 2 we briefly explain what {\it TTC} does when simplifying index tensor polynomials in general. Also we describe some {\it TTC} functions usually used in operations over this kind of objects . In section 3  we define exactly what kind of tensors we will find and which are the steps to reach our main result, that is to find all four index quadratic in Riemann and  divergence free tensors in any dimension and in any space-time. Next we specialize these results  for four dimensions and for vacuum space-times. We analyze  the tensors found with respect to index  symmetry properties, specially we give existence and uniquess theorems for totally symmetric tensors.

Finally Appendix A is devoted to show a sample {\it TTC} session where the C.D.Collinson calculations are reproduced using the essential {\it TTC} functions.    

\section{Computational simplification of index tensor polynomials}

As stated above, by simplification we mean to find a canonical, therefore unique, index tensor polynomial for any given monomial. In this process the following must be taken into account:
\begin{itemize}
\item Dummy indices: the output expression must be unique with respect to the freedom of renaming dummy indices.
\item Index symmetry properties of each singular tensor appearing in the monomial: since in this paper we work with Riemann tensor $R$ the following properties, and those which arise when diferentiating and contracting, must be taken into account
\begin{eqnarray}
R_{i j k l}=-R_{j i k l}=-R_{i j l k}=R_{k l i j}\nonumber \\
R_{i j k l}+R_{i l j k}+R_{i k l j}=0\nonumber\\
R_{i j k l ;m}+R_{i j m k ;l}+R_{i j l m ;k}=0\nonumber\\
R_{i j k l} \underbrace{_{;m ....;n}}_{k}=a\, R_{i j k l ;}\underbrace{_{n ....;m}}_{k}+({\rm Riemann\  terms})\underbrace{_{;...}}_{k-2}\label{riemp}
\end{eqnarray} 
where the last one refers to Ricci commuting relations. At this point it is important to set up the convention for Ricci commuting relations and the definition of Ricci tensor used by {\it TTC} which follows \cite{Misner}.

 If $v$ is a vector field then
\[
v_{i;j;k}  - v_{i;k;j}  = v_m \,R_{\,\,\,\,i\,j\,k}^m 
\]
\[
R_{ij}^{}  = R_{\,\,\,\,\,imj}^m 
\]

It is important to note the change of sign of the Ricci tensor with respect to that used in Sachs\cite{Sachs}, Collinson\cite{Collinson} and Robinson\cite{Ivor} original papers
\item Properties due to the finite dimension of space. This kind of properties arise due to the posssibility of antisymetrization over $n+1$ indices appearing in the monomial, $n$  being the dimension of space. In the next example $n=2$
\[
... + A_{i\,j\,} B_k  + ... \Rightarrow A_{[i\,j\,} B_{k]}  = 0
\]
\end{itemize}
In the following subsection we briefly explain how algorithms work.
For details we refer to \cite{XACyclicRicci,XADim}
\subsection{Dummy index problem}
Actually {\it TTC} uses a codification of each monomial, which is an improved version of \cite{XA1998}, following the criteria:

{\bf 1)}Free indices are renamed:\verb" IndexF[1], IndexF[2],..." all of them in upper position ( contravariant)

{\bf 2)} Dummy indices are renamed \verb" IndexD[1], IndexD[2],..." all of them in upper position ( contravariant)  

{\bf 3)} We find the permutation of free and dummy indices in order to find the most lexicographically ordered monomial expression. Note that at this stage the codification is  independent of the order of free indices. We call this {\it globalcodification}

Finally we take the invers of permutation performed on free indices over the {\it globalcodification} of the monomial. The final result is called {\it codification} of the original monomial. 

All rules found through a calculation are stored in {\it codification} version but all monomials with the same   {\it globalcodification} version can use a permutation version of the stored rules, so the work of finding rules is essentially independent of the order of free indices.

\subsection{Index tensor properties}

First the user must declare each tensor entering the monomial, together with their symetries. Some properties are automatically applied as Ricci commuting relations. 

Key functions: \verb"InputTensor, InputSymmetries"

{\it TTC} acts on each monomial applying in all possible ways the individual properties of each tensor over the starting monomial and over each new resulting monomial. When this loop ends ( no new monomials are created) {\it TTC} solves the resulting system of equations and applies the rules over the original monomial. Subsequent monomials can use the same or a permuted version of these  rules if the {\it globalcodification} version is the same as one of the monomials in the rules.   

Key functions: \verb" Index[gname,SimplifyAllIndex[2]]"

\subsection{Dimensional properties}

As stated in the introduction, dimensional propeties arise due to the possibility of antisymmetrizing $n+1$ indices of the starting monomial and equating the result to zero, $n$ being the space dimension. Nevertheless we must be sure to take all possible sets of $n+1$ indices in the monomial, even if they are dummy or {\it hidden} ( as in the curvature $R$). The algorithm takes the monomial and writes it in a metric expanded version. The possible sets of $n+1$ indices are in the metric part of the monomial.  Let us see some examples in four dimensions.

{\bf 1)} In the first example there are no dimensional properties:	
\[
R \to R_{mnpq}^{} \,g_{}^{mp} \,g_{}^{nq} 
\]

{\bf 2)}Only one set of $5$ indices can be effectively antisymmetrized:

\[
R\,R^{ab}_{}  \to R_{mrns}^{} R_{opqr}^{}  \,g_{}^{mn} \,g_{}^{ar} g_{}^{bs} g_{}^{oq} g_{}^{pr} 
\]

{\bf 3)}The last example  is the case handled in the present paper. Several sets of 5 indices can be efectively antisymmetrized. Note that for $n \ge 6$ there are no dimensional properties.
\[
R\,R_{}^{abcd}  \to R_{mnop}^{} R_{rstv}^{} \,g_{}^{mo} \,g_{}^{np} g_{}^{ar} g_{}^{bs} g_{}^{ct} g_{}^{dv} 
\]

{\it TTC} finds all these properties and simplify them using nondimensional ones. Then it solves the resulting system of equations and applies the resulting rules  when necessary. Although the procedure described is correct actually the algorithm has been improved in order to optimize the number of sets of indices to be antisymmetrized \cite{XADim}

Key functions: \verb" Index[gname,SimplifyAllIndex[3]]"

\section{Collinson like calculus}
C.D Collinson found {\it by hand } 10 independent divergence free tensors having four indices and quadratic in Riemann. Here we use {\it TTC} to make the computerized version of his calculations, but the plan is close to the Collinson paper. 

First we must generate the starting basis 
\[
T^{abcd}  = A^{abcdefghklmn} \,R_{efgh} \,R_{klmn} \, + \,B^{abcdklmnpq} \,R_{klmn;pq} 
\]
where $ A^{abcdefghklmn} \, $ and $ B^{abcdklmnpq}$ are polynomic tensors built with the metric $g^{ab}$ using the free indices plus the contracted ones in all possibles ways, symbollically
\[
\begin{array}{l}
 A^{abcdefghklmn}  = \sum\limits_\sigma  {\alpha _\sigma  \,\sigma \left( {g^{ab} g^{cd} g^{ef} g^{gh} g^{kl} g^{mn} } \right)}  \\ 
 B^{abcdklmnpq}  = \sum\limits_\sigma  {\beta _\sigma  \,\sigma \left( {g^{ab} g^{cd} g^{kl} g^{mn} g^{pq} } \right)}  \\ 
 \end{array}
\]
$\sigma$ being  the permutations over metric indices and $\alpha_\sigma$ and $\beta_\sigma$ constant parameters.

{\it TTC} performs this calculation simplifying the result using nondimensional properties and dimensional ones. In the second case we obtain, as in Collinson's paper, 63 independent monomials. This polynomial is what we call the starting basis $T$, which has 63 independents coeficients (including the global one). In doing this {\it TTC} finds 3 independent dimensional properties which includes those found by H.A.Buchdahl\cite{Buchdahl}.

Second we take the divergence of $T$
\[
T_{\,\,\,\,\,\,\,\,\,\,\,\,\,\,;a\,\,}^{abcd} 
\]
{\it TTC} calculates this and simplifies it by using nondimensional and dimensional properties.
After this we must solve the coefficients of the equation 
\begin{equation}
T_{\,\,\,\,\,\,\,\,\,\,\,\,\,\,;a\,\,}^{abcd}=0 
\label{tdivfree}
\end{equation}
so we find the family of polynomials $T$ fulfilling (\ref{tdivfree}). This computation has been done in a generic space-time and in vacuum. The results have been analyzed under a few relevant index symmetry properties. From the computations we have: 

\subsection{Without-dimension properties}

\begin{theorem}\label{gennond1} In a generic space-time and not using dimensional properties:

{\bf 1)} There exist 14 independent quadratic in Riemann four index divergence free tensors.

{\bf 2)}There are no totally symmetric tensors fulfilling 1) 

{\bf 3)} The complete family of tensors $T^{abcd}$ fulfilling  1) totally symmetric in $(bcd)$ is 
\[
\begin{array}{l}
 T_{}^{abcd}  = a_S \,T_S^{abcd}  + a_R T_R^{abcd} ; \\ 
  \\ 
 T_S^{abcd}  = Q_{}^{a(bcd)} ; \\ 
 Q_{}^{abcd}  =  - \frac{1}{3}g_{}^{ac} \,R_{\,\,\,i}^d \,R_{}^{ib} \, + \,2\,R_{\,\,\,}^{ab;cd} \, - \frac{4}{3}\,R_{\,\,\,}^{bd;ac}  +  \\ 
 \,\,\,\,\,\,\,\,\,\,\,\,\,\,\,\frac{4}{3}g_{}^{ac} R_{\,\,\,\,\,\,\,\,\,\,\,i}^{bd;i}  - \,2\,g_{}^{ac} R_{\,\,\,\,\,i}^{b\,\,\,\,\,;d\,i}  + \frac{4}{3}\,\,R_{\,\,\,i}^b \,R_{}^{acid}  +  \\ 
 \,\,\,\,\,\,\,\,\,\,\,\,\,\,\,\,2\,R_{}^{aibj} \,R_{\,\,\,i\,\,\,\,\,\,j}^{c\,\,\,\,d}  - \frac{1}{2}\,g_{}^{ac} \,R_{ij\,\,\,\,k}^{\,\,\,d} \,R_{}^{ijbk}  \\ 
  \\ 
 T_R^{abcd}  = \,X_{}^{ab} \,g_{}^{cd} ; \\ 
 X_{}^{ab}  = K\,U_{}^{ab} \, + \,L\,V_{}^{ab}  - \frac{1}{4}W_{}^{ab}  \\ 
 U_{}^{ab} \, = G_{\,\,\,\,\,\,\,\,\,\,\,s}^{ab;s}  - 2\,G_{\,\,\,\,\,\,\,\,\,\,\,s}^{sb;a} \, + \,2\,G_{\,\,\,\,\,p\,}^a R_{}^{pb}  - \frac{1}{2}\,G_{\,\,\,\,\,pq\,} R_{}^{pq} g_{}^{ab}  \\ 
 V_{}^{ab}  = \,R_{}^{;ab}  - R_{\,\,\,\,\,\,s}^{;s} \,g_{}^{ab} \, - \,R\,S_{}^{ab}  \\ 
 W_{}^{ab}  = G_{}^{apqr} \,R_{\,\,\,\,pqr}^b  - \frac{1}{4}\,g_{}^{ab} \,G_{}^{mpqr} R_{mpqr}^{}  \\ 
 G_{\,\,\,\,\,\,cd}^{ab}  = R_{\,\,\,\,\,\,cd}^{ab}  - \,4\,g_{\,\,\,\,\,\,[c}^{[a} S_{\,\,\,\,\,\,d]}^{b]}  \\ 
 S_{\,\,\,\,\,}^{ab}  = R_{\,\,\,\,\,}^{ab}  - \frac{1}{4}g_{\,\,\,\,\,}^{ab} R \\ 
 \end{array}
\]
where $a_S,a_R,K,L$ are four independent constant parameters.
\end{theorem}

$T_S$ is the  tensor found by Sachs\cite{Sachs} with an opposite sign on the term $\frac{4}{3}\,\,R_{\,\,\,i}^b \,R_{}^{acid}$ with respect to his original definition. $T_R$ is the family defined by Robinson\cite{Ivor}.

\begin{theorem} In a vacuum $R_{ij}=0$ space-time and not using dimensional properties: 

{\bf 1)} There exist 6 independent quadratic in Riemann four index divergence free tensors.

{\bf 2)} There is only one tensor $T^{abcd}$ fulfilling 1) and totally symmetric:
\[
\begin{array}{l}
 T^{abcd}  = M^{(abcd)}  + 16\,P^{a(bcd)}  - 4\,Q^{(abc)d}  - 4\,Q^{(a|d|bc)} ; \\ 
 M^{abcd}  = g^{ab} \,g^{cd} \,R_{mnop} R^{mnop} ; \\ 
 P^{abcd}  = R_{m\,\,\,\,n}^{\,\,\,a\,\,\,\,\,\,\,b} \,\,R^{mcnd} ; \\ 
 Q^{abcd}  = g^{ab} \,\,R_{mno}^{\,\,\,\,\,\,\,\,\,\,\,c} \,\,R^{mnod}  \\ 
 \end{array}
\] 
\end{theorem}

\subsection{Using dimensional properties}

\begin{theorem}[Collinson corrected]\label{gend1} In a generic four dimensional space-time:

{\bf 1)} There exist 9 independent quadratic in Riemann four index divergence free tensors.

Using the same notation of Collinson \cite{Collinson} the Collinson family is
\[
\begin{array}{l}
T^{abcd}  = a_1 \,T_1^{abcd}  + a_2 \,T_2^{abcd}  + a_3 \,T_3^{abcd}  + a_4 \,T_4^{abcd}  + a_5 \,T_5^{abcd}  + a_6 \,T_6^{abcd}  +\\ 
\, \, \, a_8 \,T_8^{abcd}  + a_9 \,T_9^{abcd}  + a_{10} \,T_{10}^{abcd} 
\end{array}
\]
where $a_i$ are arbitrary numerical coefficients. 

The tensors $T_1$ to $T_9$ (without $T_7$) can be writen in terms of $T_{10}$
\[
\begin{array}{l}
 T_1^{abcd}  = T_2^{acbd}  = T_3^{adcb}  = g^{cd} T_{10\,\,\,\,i}^{a\,i\,\,\,b}  \\ 
 T_6^{abcd}  = T_5^{acbd}  = T_4^{adcb}  = g^{cd} T_{10\,\,\,\,i}^{a\,b\,\,\,\,i}  - \frac{1}{2}g^{cd} T_{10\,\,\,\,i}^{a\,i\,\,\,b}  \\ 
 3\,T_8^{abcd}  = T_{10}^{acbd}  - T_{10}^{abcd}  \\ 
 \,T_9^{abcd}  = T_{10}^{adcb}  - 2\,T_{10}^{abcd}  \\ 
 \end{array}
\]
and $T_{10}$ is defined through $
T_{10}^{abcd}  = 6\,Q_1^{abcd}  + Q_2^{ab(cd)}  + Q_3^{ab(cd)}$ being
\[
\begin{array}{l}
 Q_1^{abcd}  = R_{\,\,\,i}^c \,R_{\,\,\,}^{bdai}  + R_{\,\,\,i}^b \,R_{\,\,\,}^{daci}  - R_{\,\,\,\,\,\,\,\,\,\,\,;i}^{aibc\,\,\,\,\,;d}  + R_{\,\,\,\,\,\,\,\,\,\,\,;i}^{acdi\,\,\,\,;b}  + g_{\,\,\,}^{ab} R_{\,\,\,\,\,\,\,\,\,\,;ij}^{dicj}  \\ 
 Q_2^{abcd}  = 4\,R_{\,\,\,\,\,\,\,\,\,\,\,\,;i}^{aibd\,\,\,\,\,\,\,c}  - 6\,g_{\,\,\,}^{ad} R_{\,\,\,\,\,\,\,\,\,;i}^{ib\,;c}  - 6\,g_{\,\,\,}^{ad} R_{\,\,\,\,\,\,\,\,\,;ij}^{icbj\,}  + 6\,R_{}^{ab;cd}  + 6\,R_{\,\,\,\,\,\,\,\,\,\,;i}^{abci\,\,\,;d}  \\ 

 Q_2^{abcd}  = 8\,g_{\,\,\,}^{ac} R_{ij\,\,\,k}^{\,\,\,\,d} \,R_{\,\,\,\,\,\,\,\,}^{jkbi}  - 8R_{\,i\,\,\,\,\,\,j\,}^{\,\,cd} R_{}^{biaj}  + 8R_{\,i}^{\,\,d} R_{}^{bica}  - 8R_{\,i\,\,\,\,\,\,j\,}^{\,\,bd} R_{}^{ciaj}  + \\ \ \ \ \ \ \ \ \ \ \ \ 2R_{\,\,\,\,i}^d R_{}^{baic}  - 2R_{\,i\,\,\,\,\,\,j\,}^{\,\,db} R_{}^{ciaj}  - 3g_{}^{ab} R_{\,\,\,i}^d R_{}^{ci}  + 3g_{}^{ab} R_{ij\,\,\,\,\,\,k}^{\,\,\,\,\,\,d} \,R_{}^{jkci}  \\ 
 \end{array}
\]

{\bf 2)}There are no totally symmetric tensors fulfilling 1) 
\end{theorem}

We note that the Collinson tensor $T_7$ does not appear, which is identically zero, since $T_{10}$ is in fact symmetric with respect to the last two indices (this is not mentioned in Collinson's paper).

\begin{theorem} In a vacuum $R_{ij}=0$ four dimensional space-time:

{\bf 1)} There exists 1 quadratic in Riemann four index  divergence free tensor.

{\bf 2)} The tensor found in 1) is totally symmetric and is the Bel-Robinson tensor \cite{BelRobinson}.

\end{theorem}

\section{Concluding remarks}

Trough four theorems we have reviewed and improved some results on classical gravitational superenergy tensors.
The results have been reached using computational algorithms which do automatic calculations. As we have commented in the introduction some trust problems could arise.  

Nevertheless we think that trust like problems are not excusively related to computational techniques. How can we be sure that {\it all} calculation of C.D. Collinson, or anyone, are correct ?. Collinson is a good pioneer on large calculations on the topic but to validate his results we hope some other should obtain the same results using independent calculations. We also hope that any other program will be able to do our calculations with an independent code and/or algorithms... 

Assumming that these computational techniques are validated and limiting our scope to superenergy tensors we think that they are a very good complement to research based on defining such tensors having some choosen properties. We can analyze the uniqueness of these definitions with respect to the desired properties. But we can also analize more complex problems, specially those related to the existence of conserved quantities. These studies can be done on a generic space- time and using Einstein field equations. We hope that in the near future we will be able to handle the case of cosmological vacuum and scalar field interaction.

\section*{Acknowledgments}
We want to kindly thank J.M.M.Senovilla to direct our work on superenergy tensors.
X.J. would like to thank the Comisi\' on Asesora de Investigaci\' on Cient\' \i fica y T\' ecnica for partial financial support, under Contract No. PB96-0384. The authors would also like to thank the Laboratori de F\' \i sica Matem\` atica at the Societat Catalana de F\' \i sica for partial financial support.

\section*{Appendix: The {\it TTC} master session }
This appendix is devoted to show a sample of {\it TTC} session so that somebody else can reproduce the calculations using{\it TTC}. In this session we find the result of Theorem(\ref{gend1}).
The inputs are indicated by \verb"In[]:=". The outputs have no special sign. The comments are enclosed by \verb"(* comments *)".

{\fontsize{8}{8}
\begin{verbatim}

(* first we load the program ttc.m. we load also eines.m 
where some private tools are defined.*)
In[]:
<<ttc.m
<<eines.m
 ----------------------------------------
 |Tools  of  Tensor  Calculus 4.1.0     |
 |  A.Balfagon, P.Castellvi and X.Jaen  |
 |     http://baldufa.upc.es/ttc        |
 |     e-mail: ttc@baldufa.upc.es       |
 |     version: september,XX,1999       |
 ----------------------------------------
 |     Session started on               |
 |     September, XX, 1999              |
 |     at 16 h 8  min 52 s              |
 ----------------------------------------
 CPU Time=0. s Memory in use=2.434 Mb 
 ----------------------------------------
 |       Eines Personals 1.0            |
 |       X.Jaen & A.Balfagon            |
 |     version: juny, 23, 1999          |
 ----------------------------------------
 (* Here we declare the name of the coordinates and dimension to be used when needed. We also introduce a
metric and a Riemann tensor.
Finally we introduce the indices we want to see in outputs *)

In[]:=(
InputCoordinates[cx4,4];
ScalarBasis=cx4;ScalarBasisQ=False;
InputSMetric[gn,cx4,"g",g];
InputSRiemann[gn,cx4,"R",Rie,Ric,R];
InputIndex[{i,j,k,l,m,n,o,p,q}]);

(*This is usefull if we want to store in "filename" all rules 
generated by TTC. If you use it remember end save using
EndSimplifyAllIndexSave*)

(*IniSimplifyAllIndexSave[filename]*)

(* generator is where the kind of tensors we study are defined*)

In[]:=generator=
Rie[a,b,c,d] Rie[-a,-b,e,f]+Rie[a,c,d,e,.;-a,.;f]//Index[gn]

 i j k l         k l   m n i j     m i j k     .;l 
0         + R         R         + R
             m n                           .;m
(* IndexTensorBasis takes the generator and uses some easy tricks
to build the starting basis, simplifying them using
 nondimensional and dimensional properties. It uses A[number]
 as parameters. The result is stored as base*)
 
In[]:=generator//Index[gn,IndexTensorBasis[3,A,base,CoRiemannRules[gn,cx4]]];

(* see *)
(*EndSimplifyAllIndexSave*)
(* we can see the starting basis)
In[]:=base
 
 i j k l          .;k .;l   i j          .;j .;l   i k 
0         + A[1] R         g     + A[2] R         g     +......

           i k   j l             m n o p           i j   k l             m n o p 
>   A[62] g     g     R         R         + A[63] g     g     R         R
                       m n o p                                 m n o p


(* Here we define A[1].. A[63] as constants parameters*)
In[]:=SetConstantCoeff[A,cx4,63];

(* Here we compute the divergence of base and simplify up to
 dimensional properties assigning the result to dbase *)

In[]:=dbase=base[.;-i]//Index[gn,SimplifyAllIndex[3]];

(* Here we find the parameters which solves dbase=0 assigning
 the result to rules*) 

In[]:=rules=SolveCoeff[dbase,A][[1]];

(*Here we apply rules to the starting basis base assigning
 the result to baseXA*)

In[]:=baseXA=base/. rules;

(* baseXA fulfill the condition imposed...*)
In[]:=baseXA[.;-i]//Index[gn,SimplifyAllIndex[3]]
0

(* Here we find t1xa ...t9xa which are the nine independent tensors
 we have found corresponding to the nine independent coefficients A[number]
appearing in baseXA*)

In[]:=t1xa=baseXA/.
 {A[6]->1,A[9]->0,A[12]->0,A[58]->0,A[59]->0,
  A[60]->0,A[61]->0,A[62]->0,A[63]->0};
 
In[]:=t2xa=baseXA/.
 {A[6]->0,A[9]->1,A[12]->0,A[58]->0,A[59]->0,
  A[60]->0,A[61]->0,A[62]->0,A[63]->0};

In[]:=t3xa=baseXA/.
  {A[6]->0,A[9]->0,A[12]->1,A[58]->0,A[59]->0,
   A[60]->0,A[61]->0,A[62]->0,A[63]->0};

(* also for tx4,tx5,tx6,tx7,tx8*)

In[]:=t9xa=baseXA/. 
  {A[6]->0,A[9]->0,A[12]->0,A[58]->0,A[59]->0,
   A[60]->0,A[61]->0,A[62]->0,A[63]->1};

(* we can see some of them*)
In[]:=t4xa 

 i j k l          2  i l   j k          2  i k   j l              j l   i k 
0         +      R  g     g     -      R  g     g     + 2      R g     R     - 

 
            j k   i l              i l   j k              i k   j l            j l     i   m k 
  2      R g     R     - 2      R g     R     + 2      R g     R     - 2      g     R     R     + 
                                                                                     m 
 
          i l     j   m k            j k     i   m l            i k     j   m l 
  4      g     R     R     + 2      g     R     R     - 4      g     R     R     - 
                m                          m                          m 
 
          i l   j k         m n            i k   j l         m n            i k .;j .;l 
  2      g     g     R     R     + 2      g     g     R     R     - 2      R             + 
                      m n                              m n 
 
          i l .;j .;k            j k .;i .;l            i l   j k     .;m 
  2      R             + 2      R             - 2      g     R             - 
                                                                  .;m 
 
          j l .;i .;k            i k   j l     .;m              l   m i j k 
  2      R             + 2      g     R             - 2      R     R         + 
                                           .;m                m 
 
            k   m i j l              j   m i k l              i   l   m j n k 
  2      R     R         + 2      R     R         + 2      R         R         - 
          m                        m                        m   n 
 
            i   k   m j n l              i l   m n j k              i k   m n j l 
  2      R         R         -      R         R         +      R         R         + 
          m   n                      m n                        m n 
 
            i j   m n k l          j l         i   m n o k          j k         i   m n o l 
       R         R         -      g     R         R         +      g     R         R
        m n                              m n o                            m n o


(* now we supose that we have find baseC which is
the baseXA expressed ussing the basis found by Collinson.
We use B[1]..(B[7])...B[10] to name the parameters ( without B[7])
accordingly to Collinson notation. We take baseC and use 
IndexUpdate to define T in such a manner that we can use T
as a tensor in their own right. This will permit us
to study symmetry index*)

In[]:=IndexUpdate[T,":=",baseC]

(*Here we study the symmetry between the first two indices, 
simplifying them up to dimensional properties and finding 
the condition for the parameters B[n]*) 

In[]:=T[a,b,c,d]-T[b,a,c,d]//Index[gn,SimplifyAllIndex[3]];
 
In[]:=rules=SolveCoeff[%,B][[1]]
 
{B[4] -> 0, B[5] -> 0, B[2] -> 0, B[3] -> 0, B[8] -> 0, B[9] -> 0, B[10] -> 0}

(* Only  B[1] and B[6] are independent.
It is in this way how we have analized
symmetry properties in this paper*)

\end{verbatim}
}

\end{document}